\documentclass[a4paper]{article} 
\pdfoutput=1
\usepackage[english]{babel}
\usepackage[pdftex]{graphicx} 
\usepackage[bf,small,nooneline]{caption}
\usepackage{citesupernumber}

\author{M.\ Holmstr\"{o}m\thanks{
  Swedish Institute of Space Physics, PO~Box~812, SE-98128~Kiruna, Sweden.}, 
A.\ Ekenb\"{a}ck\footnotemark[1], 
F.\ Selsis\thanks{Centre de Recherche Astrophysique de Lyon (CNRS, Universit\'e de Lyon, Ecole Normale Sup\'erieure de Lyon), 46 All\'ee d'Italie, F-69007, Lyon, France.} \hspace{0cm}\thanks{Laboratoire d'Astrophysique de Bordeaux (CNRS, Universit\'e Bordeaux 1), BP 89, F-33270, Floirac, France.}, 
T.\ Penz\thanks{INAF -- Osservatorio Astronomico di Palermo, 
  Piazza del Parlamento 1, I-90134 Palermo, Italy.}, 
H.\ Lammer\thanks{Space Research Institute, Austrian Academy of Sciences, 
  Schmiedlstr.\ 6, A-8042, Graz, Austria.}, and 
P.\ Wurz\thanks{Physikalisches Institut, University of Bern, Sidlerstr. 5, 
  CH-3012 Bern, Switzerland.} 
}
\title{\textbf{\textsf{
Energetic neutral atoms as the explanation for the
high velocity hydrogen around HD~209458b
}}}
\date{December 7, 2007}

\begin{document}
\maketitle

\textbf{
Absorption in the stellar Lyman-$\alpha$ (Ly-$\alpha$) line observed during 
the transit of the extrasolar planet HD~209458b reveals high velocity atomic 
hydrogen at great distances from the planet\cite{Vidal-Madjar03,Ben07}. 
This has been interpreted as hydrogen atoms escaping from the exosphere 
of the planet\cite{Vidal-Madjar03,Vidal-Madjar03b}, 
possibly undergoing hydrodynamic blow-off\cite{Vidal-Madjar04}, 
being accelerated by stellar radiation pressure.  
However, around solar system planets the production of energetic neutral 
atoms from charge exchange between solar wind protons 
and neutral hydrogen from the exospheres 
has been observed\cite{Collier01,FutaanaJet06,GalliVEX07}, 
and should also occur at extrasolar planets. 
Here we show that the measured transit-associated Ly-$\alpha$ absorption 
can be explained by the interaction between the exosphere of HD~209458b 
and the stellar wind, and that radiation pressure alone cannot explain 
the observation.  
This is the first observation of energetic neutral atoms outside 
the solar system. 
Since the stellar wind protons are the source of the observed 
energetic neutral atoms, this provides a completely new method 
of probing stellar wind conditions, and our model suggests a 
slow and hot stellar wind near HD~209458b at the time of the observation. 
} 

Energetic neutral atoms (ENAs) 
are produced wherever energetic ions meet a neutral atmosphere, 
and solar wind ENAs have been observed at every planet in the 
solar system where ENA instrumentation has been available --- 
at Earth\cite{Collier01}, at Mars\cite{FutaanaJet06}, and 
at Venus\cite{GalliVEX07}. 

By energetic we mean that the ions have a much greater velocity than the 
thermal velocities of the exospheric neutrals. 
During the charge exchange process, an 
electron is transferred from the neutral to the ion, resulting 
in a neutral atom and an ionized neutral.  
Due to the large relative velocities of the ions and the exospheric 
neutrals, the momenta of the individual atoms are 
preserved to a good approximation.  Thus, the produced ENAs
will have the same velocity distribution as the source population of ions.

When first observed (also by their Ly-$\alpha$ 
signature\cite{Barth68,Barth69}), the extended hydrogen coronae of Mars 
and Venus were assumed to constitute the uppermost layers of an escaping 
exosphere. The observed densities were used to infer exospheric scale 
heights and temperatures, which proved to be extremely high compared to 
theoretical predictions (up to 700~K). 
In situ spacecraft observations later found exospheric temperatures 
of $\sim$210 and $\sim$270~K\cite{Lichtenegger06}. 
The discrepancy was eventually explained by 
photochemically-produced energetic particles, and by 
ENAs, produced by charge exchange between energetic solar wind protons and 
the planetary exosphere. This mechanism, well-known in the solar system, 
has however not been considered as a possible origin of the atomic hydrogen 
corona revealed by HST observations of HD~209458b. 

HD~209458b is a Jupiter-type gas giant with a mass of 
$\sim$0.65~M$_\mathrm{Jup}$ 
and a size of $\sim$1.32~R$_\mathrm{Jup}$ that orbits 
around its host star HD~209458 at $\sim$0.045~AU\cite{Knutson07}, 
which is a solar-like G-type star with an age of about 4~Gyr.
The activity of the star can be estimated from its X-ray 
luminosity measured by the XMM-Newton space observatory, and is 
comparable to that of the present Sun during a moderately quiet 
phase\cite{penz07b}.  Because of its Sun-like stellar type and average 
activity, it is justified to use the energy 
environment observed at the Sun as inputs for our model.

For a first estimate of the ENA production near HD~209458b, we 
assume that the charge exchange takes place in an undisturbed 
stellar wind that is flowing radially away from the star. 
At 0.045~AU from HD~209458, the stellar wind is most likely 
subsonic\cite{preusse05} and does not produce a planetary bow shock. 
Simulations indicate a subsolar magnetopause distance of about 
four planetary radii if the planet is magnetized\cite{preusse06}. 
If the planet is not magnetized, we would expect the undisturbed 
stellar wind to get even closer to the planet. 
Here we model the ENA production by a particle model that includes 
stellar wind protons and atomic hydrogen. 
Charge exchange between stellar wind protons and exospheric hydrogen 
atoms takes place outside a conic obstacle that represents the 
magnetosphere of the planet (Supplementary Fig.~1). 
The resulting exospheric cloud, along with the produced ENAs, 
covers a region larger than the stellar disc, 
as seen from Earth~(Fig.~\ref{fig:cloud}). 
The cloud is shaped like a comet tail due to the stellar radiation pressure, 
curved by the Coriolis force, 
as predicted\cite{Schneider98} and seen in earlier 
numerical simulations\cite{Vidal-Madjar03,Vidal-Madjar03b}. 
There is a population of atoms with high velocity 
 --- these are the stellar wind protons that have 
charge exchanged, becoming ENAs. 
In the velocity spectrum along the $x$-axis (the planet--star line), 
the ENAs are clearly visible 
as a distribution that is separate from the main exospheric 
hydrogen component, due to the different bulk velocities 
and temperatures~(Fig.~\ref{fig:vel}). 

Now we estimate how the ENA cloud would affect the observed Ly-$\alpha$ 
absorption spectrum of HD~209458b\cite{Vidal-Madjar03}. 
The line profile was observed outside and during transit, and the 
difference between the two profiles correspond to the attenuation by 
hydrogen atoms~(Fig.~\ref{fig:cmp}). 

There are several features of the transit spectrum 
that any proposed source of the observed hydrogen atoms need to account for. 
First, hydrogen atoms with velocities of up to -130~km/s (away from the star). 
Second, a fairly uniform absorption over the whole velocity range 
-130 to -45~km/s. 
Third, absorption in the velocity range between 30 and 105~km/s 
(toward the star). 

The current explanation of the observation is that hydrogen atoms in 
the exosphere are undergoing hydrodynamic escape, and are then further 
accelerated by the stellar radiation 
pressure\cite{Vidal-Madjar03,Vidal-Madjar04}.  
There are however some difficulties in explaining the observations by this 
process, as can be seen by examining the three features listed above. 

First of all, a large radiation pressure on the hydrogen atoms is needed 
to accelerate them to a velocity of 130~km/s before they are 
photoionized.  The acceleration also has to occur before they 
move out from the region in front of the star, 
due to the orbital motion of the planet. 
The second feature is difficult to explain, since if hydrogen 
atoms were driven to speeds of up to 130~km/s, we would expect the 
velocity spectrum to have an exponential decay for higher velocities, 
due to the finite lifetime of hydrogen atoms because of photoionization 
(four hours on average).  This drop-off for high 
velocities is independent of the details of the model, e.g., the values 
of radiation pressure and photoionization lifetime used. 
This would lead to a decay in the absorption spectrum, inconsistent 
with the observed fairly uniform absorption over the whole velocity range 
-130 to -45~km/s.  
Finally, an exosphere driven by radiation pressure cannot explain 
hydrogen atoms moving toward the star with speeds between 30 and 105~km/s. 
However, this feature of the observation is not completely certain, and 
it has been stated\cite{Vidal-Madjar03} 
that more observations are needed to clarify whether an
absorption is present in the red part of the line (toward the star).

Our model shows that all observed features listed above can be explained by 
ENAs.  If we turn off the ENA production in the model, none of these features 
are explained.  
When we compare the modeled Ly-$\alpha$ profile 
with the observed ones, 
we find that the modeled spectrum leads to attenuation over 
the whole velocity range from -130 to -45~km/s, as is seen in the observation. 
The model also shows some absorption in the red part of the velocity 
spectrum, i.e.\ hydrogen atoms moving at high velocities toward the star, 
since for this stellar wind (50~km/s and $10^6$~K), some 
part of the proton velocity distribution will have positive velocities 
along the $x$-axis (toward the star), 
resulting in an ENA flux toward the star. 
This slow and hot stellar wind is not unrealistic at such 
small orbital distances\cite{preusse05}. 

If ENAs from charge exchange are responsible for the observed attenuation, 
we have on one hand, much less information on the main exospheric component 
than suggested by previous explanations. 
Since what we observe are mainly ENAs, a range of 
exospheric conditions and atmospheric loss rates 
can be consistent with the observation. 
Thus, the observation only constrains the radiation pressure driven 
atmospheric escape insofar that the exosphere has to be extended enough 
to reach the stellar wind outside the magnetopause of the planet. 
The atmospheric escape through ENA production is small. 
For the model parameters used here, the loss is $7\cdot10^5$~kg/s, which is 
more than an order of magnitude smaller than the estimated thermal loss 
of about $10^7$~kg/s for similar exospheric 
conditions\cite{Yelle04,Yelle06,Munoz07}. 

On the other hand, we gain information on the underlying plasma 
flows, and if it is the undisturbed stellar wind, we have a way of 
observing stellar wind properties such as temperature and velocity
around other stars, at the location of extrasolar planets. 
By varying the stellar wind temperature, the stellar wind velocity 
and the radiation pressure in the model, 
we find a best fit of the modeled Ly-$\alpha$ absorption to the 
observation for a stellar wind velocity of 50~km/s, 
and a temperature of 10$^6$~K~(Fig.~\ref{fig:cmp}). 

Although, for HD~209458b, the available Ly-$\alpha$ data are affected by large 
uncertainties, more accurate observations 
would improve the derived stellar 
wind estimates.  Also, the depletion of the stellar wind proton 
flow by charge exchange will change the character of the 
planet--stellar wind interaction.  Present models of the 
stellar wind interaction with HD~209458b have not taken this process 
into account\cite{preusse06}. 
More observations of the Ly-$\alpha$ absorption by HD~209458b, 
and its variation over time, could be used to confirm the origin 
of the extended hot hydrogen cloud.  The variability of the Ly-$\alpha$ 
by ENAs should be larger on short time scales 
than for other explanations, e.g., hydrodynamic escape, since the solar 
wind parameters can change significantly on a timescale of hours, 
as seen near Earth. 


\newpage

\noindent\textbf{\textsf{Supplementary Information}} 
is linked to the online version of the 
paper at \texttt{www.nature.com/nature}
\vspace{1em}

\noindent\textbf{\textsf{Acknowledgements}}\\ 
  M.H., H.L., F.S. and T.P.\ thanks the International Space Science 
  Institute (ISSI), as this study was carried out
  within the framework of the ISSI team ``Evolution of Exoplanet Atmospheres
  and their Characterization''. 
  T.P.\ is supported by the Marie Curie Fellowship project ISHERPA, 
  and the host institution INAF-Osservatorio Astronomico di Palermo.
  H.L.\ thanks ASA for funding the CoRoT project. 
  This research was conducted using the resources of the 
  High Performance Computing Center North (HPC2N), Ume\aa\ University, Sweden, 
  and the Center for Scientific and Technical Computing (LUNARC), 
  Lund University, Sweden. 
  The software used in this work was in part developed by the 
  DOE-supported ASC / Alliance Center for Astrophysical 
  Thermonuclear Flashes at the University of Chicago.
\vspace{1em}

\noindent\textbf{\textsf{Author Contributions}}\\ 
A.E.\ wrote an initial version of the simulation code. 
F.S.\ helped in modeling the observation. 
T.P.\ provided knowledge on the atmospheres of extrasolar planets. 
H.L.\ and F.S.\ suggested that the HST observation could be due to ENAs. 
P.W.\ contributed with expertise in ENA processes. 
\vspace{1em}

\noindent\textbf{\textsf{Author Information}}\\ 
Reprints and permissions information is available at
\texttt{www.nature.com/reprints}. 
The authors declare no competing financial interests.
Correspondence and requests for materials should be addressed to M.H.\ 
(\texttt{matsh@irf.se}).

\newpage

\begin{figure*}[ht]
\caption{ 
The hydrogen cloud around the planet. 
Shown from (a) above, perpendicular to the planet's orbital plane, and 
(b) from Earth, along the $x$-axis direction. 
Each point corresponds to a hydrogen meta particle. 
The color of the points shows the velocity of the particles along the 
$x$-axis. 
Particles with velocity magnitude smaller than 50~km/s are red, 
and those with higher velocity are black. 
The small circles show the planet size.
The large circle in (b) shows the star's position at mid transit. 
During transit the star moves from left to right in (b). 
At the outer boundaries of the simulation domain, stellar wind protons 
are injected with a $2\cdot 10^3$~cm$^{-3}$ number density, 
50~km/s velocity, and $10^6$~K temperature. 
The planet's interaction with the stellar wind is modeled by removing 
all stellar wind protons inside a conic obstacle at a sub-stellar 
distance of about $4.2 R_p$ (where the radius of the planet $R_p=9.4\cdot10^7$~m). 
Hydrogen atoms are launched from an inner boundary 
(a sphere of radius $2.1 R_p$) assuming a 
number density of $10^{8}$~cm$^{-3}$ and a temperature of 
$7000$~K, consistent with atmospheric models\cite{Munoz07}. 
The trajectory of each proton and hydrogen atom is followed 
in time.  The forces on a hydrogen atom are the gravity of the planet, 
the Coriolis force due to the rotating coordinate system, and 
radiation pressure.  After each time step a hydrogen atom can undergo 
photoionization, elastic collision with another hydrogen atom or 
charge exchange with a proton. 
The photoionization time assumed is 4 hours which is a scaled Earth value. 
The radiation pressure corresponds to a photon-hydrogen collision rate 
of 0.35~s$^{-1}$ and is chosen to improve the model fit. 
It is lower than a scaled Earth value of $0.6-1.6$~s$^{-1}$ over a 
solar cycle. 
The coordinate system used is centered at the planet 
with its $x$-axis toward the star, and the $y$-axis opposite to 
the planet's velocity. 
Further details of the simulations can be found in the SI. 
        } \label{fig:cloud}
\end{figure*}

\begin{figure}[ht]
\caption{ 
Velocities of the hydrogen atoms. 
The modeled $x$-axis (planet--star) velocity spectrum of 
hydrogen atoms in front of the star at the moment of mid transit, 
not including atoms in front or behind the planet. The part of the 
distribution that is due to ENAs is shaded. 
Varying the stellar wind temperature and velocity in the model 
confirms that the width of this part of the distribution is proportional 
to the temperature of the stellar wind, with a larger width for larger 
temperatures, and the center of the distribution 
follows the stellar wind velocity. 
The un-shaded part of the spectrum is due to the exospheric hydrogen atoms. 
        } \label{fig:vel}
\end{figure}

\newpage

\begin{figure}[ht]
\caption{Comparison of the modeled Ly-$\alpha$ profile with the observed 
ones.  In blue is the observed profile before transit.  In green is the 
observed profile during transit.  In red is the modeled profile, 
constructed by applying the attenuations computed from the simulations to 
the observed profile before transit. 
The abscissa is the hydrogen velocity along the $x$-axis 
(away from Earth --- toward the star). 
The regions where there is a significant difference between the profiles are
denoted 'In' and 'Geo', the latter being the region of geocoronal 
emission at low velocities that should be excluded. 
The modeled profile is computed at the instant of mid-transit.  
The details of computing the Ly-$\alpha$ 
attenuation from the hydrogen cloud are given in the SI.  
The modeled Ly-$\alpha$ absorption shown here is for 
a stellar wind velocity of 50~km/s and a temperature of 10$^6$~K. 
The fit is worse for stellar wind velocities of 0 or 100~km/s, 
or stellar wind temperatures of $2\cdot10^6$~K or $0.5\cdot10^6$~K 
as shown in the SI. 
        } \label{fig:cmp}
\end{figure}

\newpage

Figure~\ref{fig:cloud}
\vspace{15em}

\begin{figure*}[hp]
\begin{center}
  \includegraphics[width=\columnwidth]{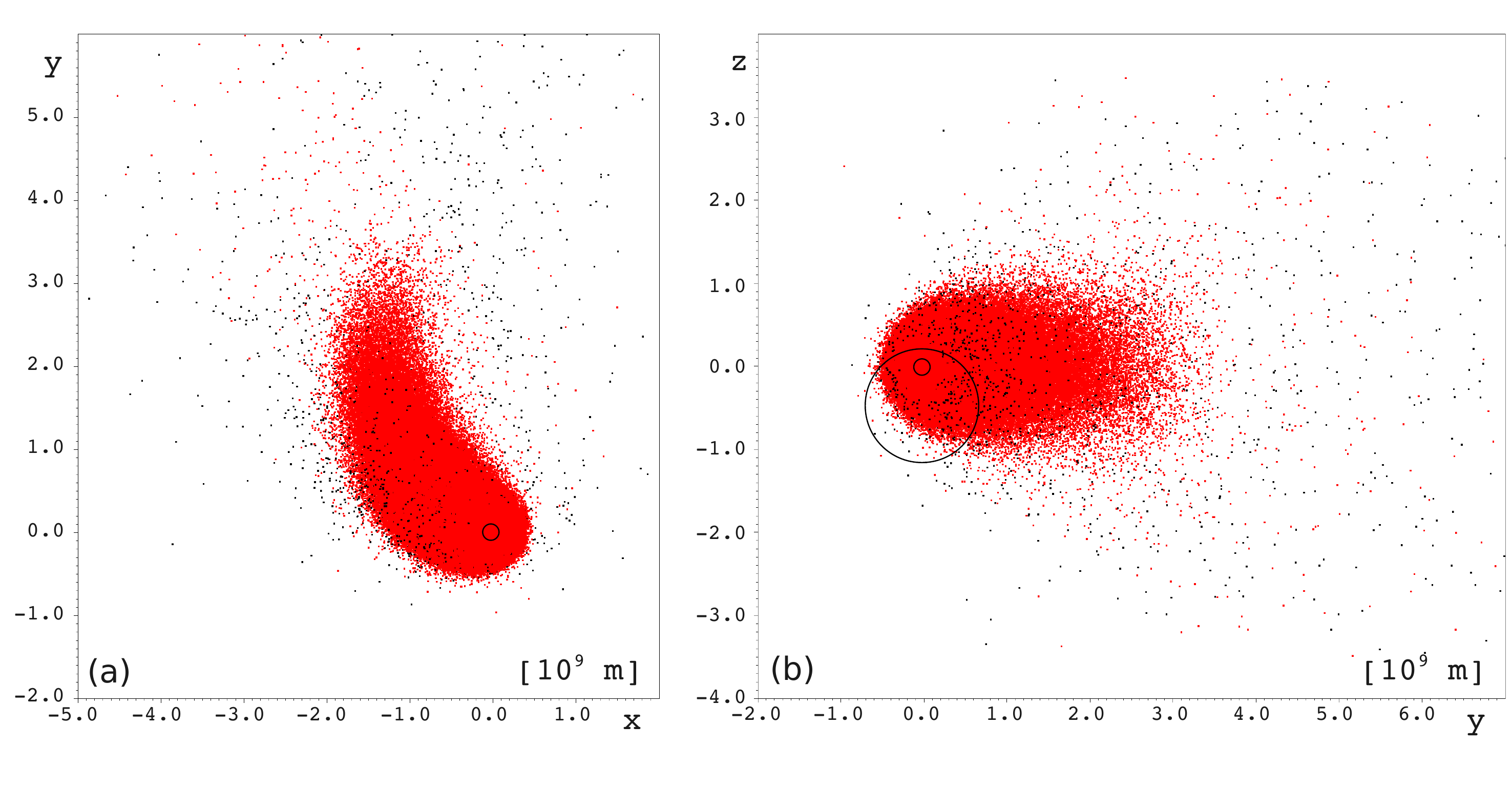}
\end{center}
\end{figure*}

\newpage

Figure~\ref{fig:vel}
\vspace{15em}

\begin{figure}[hp]
\begin{center}
  \includegraphics[width=0.8\columnwidth]{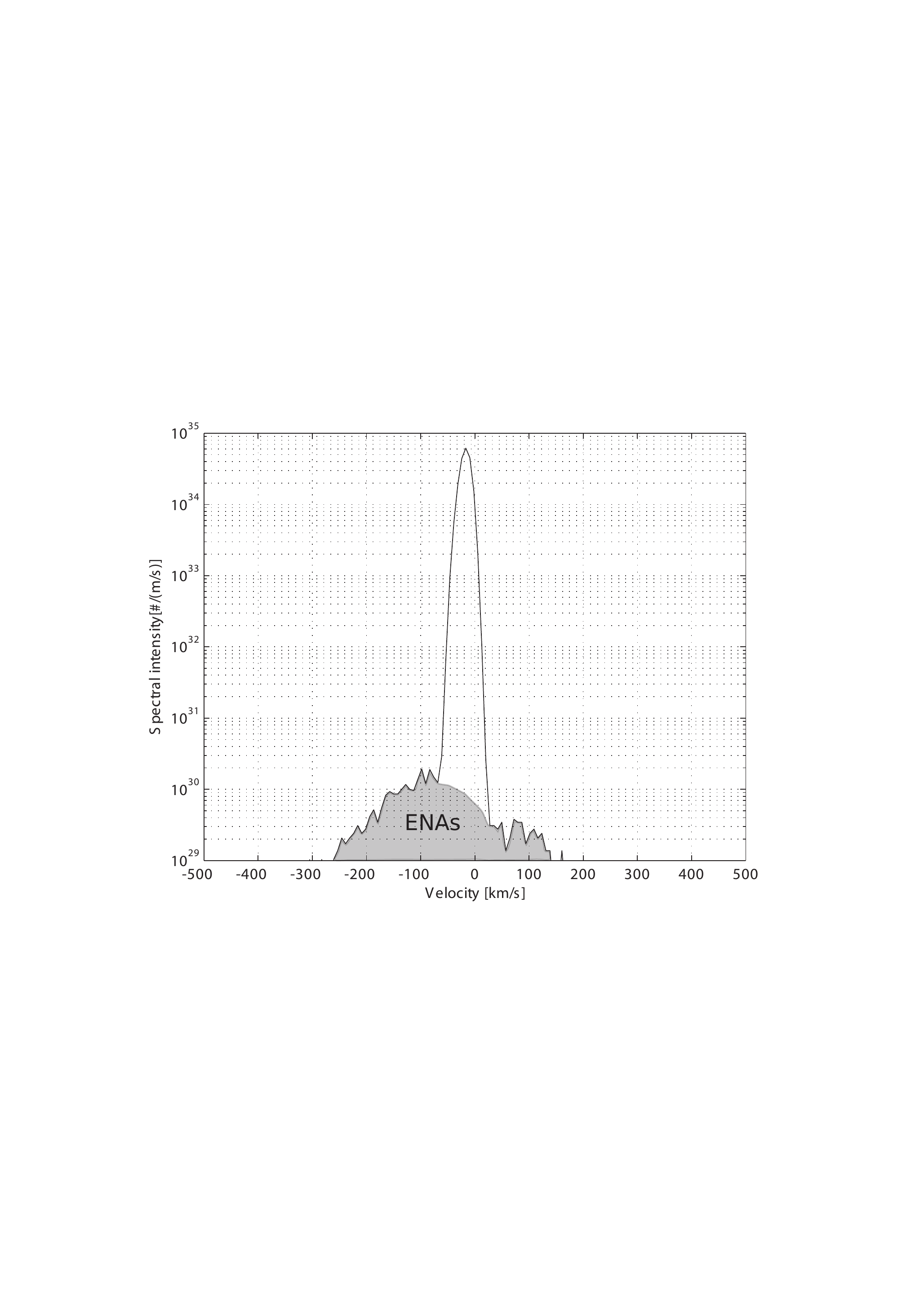}
\end{center}
\end{figure}

\newpage

Figure~\ref{fig:cmp}
\begin{figure}[hp]
 \begin{center}
  \includegraphics[width=0.8\columnwidth]{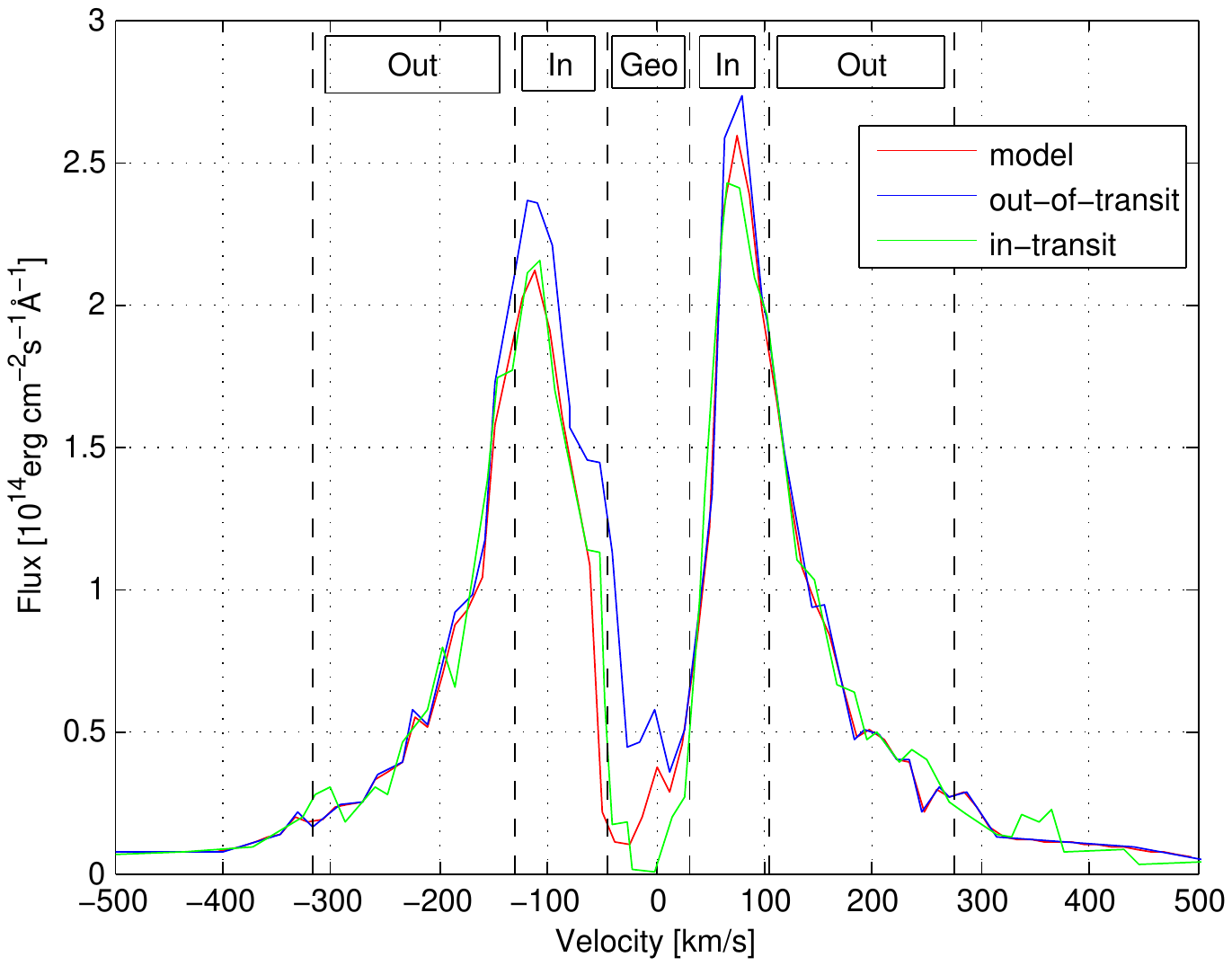}
\end{center}
\end{figure}

\end{document}